\documentclass[aps,prl,twocolumn,superscriptaddress]{revtex4-1}
\usepackage{epsfig,graphicx,amssymb,amsmath}

\DeclareMathOperator{\sech}{sech}

\begin{document}

\title{Inhomogeneous spin domain induced by the quadratic Zeeman effect in spin-1 Bose-Einstein condensate}
\author{Dun Zhao}
\affiliation{Beijing National Laboratory for Condensed Matter
Physics $\&$ Institute of Physics, Chinese Academy of Sciences, Beijing
100190, China}
\affiliation{School of Mathematics and Statistics $\&$ Center for Interdisciplinary Studies, Lanzhou 730000, China}

\author{Shu-Wei Song}
\affiliation{Beijing National Laboratory for Condensed Matter
Physics $\&$ Institute of Physics, Chinese Academy of Sciences, Beijing
100190, China}

\author{Lin Wen}
\affiliation{Beijing National Laboratory for Condensed Matter
Physics $\&$ Institute of Physics, Chinese Academy of Sciences, Beijing
100190, China}

\author{Zai-Dong Li}
\affiliation{Hebei University of Technology, Tianjin 300401, China}

\author{Hong-Gang Luo}
\affiliation{School of Mathematics and Statistics $\&$ Center for Interdisciplinary Studies, Lanzhou 730000, China}

\author{Wu-Ming Liu}
\affiliation{Beijing National Laboratory for Condensed Matter
Physics $\&$ Institute of Physics, Chinese Academy of Sciences, Beijing
100190, China}

\begin{abstract}
We show two kinds of inhomogeneous spin domain possessing N\'{e}el-like domain walls in spin-1 Bose-Einstein condensate, which are induced by the positive and negative quadratic Zeeman effect (QZE) respectively. In both cases, the spin density distribution is inhomogeneous and has zeros where the magnetization vanishes. For positive and negative QZE, the spin patterns and topological structures are remarkably different. Such phenomena are due to the pointwise different axisymmetry-breaking caused by the pointwise different population exchange between the sublevels, arising uniquely from the QZE.  We analyze in detail the inhomogeneous domain formation and related experimental observations for the spin-1 $^{87}$Rb and $^{23}$Na condensate.
\end{abstract}

\maketitle

\textit{Introduction--.}  The experimental realization of spinor Bose-Einstein condensate (BEC) in an optical trap \cite{Stamper-Kurn1998,Bloch2008} provides a unique opportunity to study the quantum magnetism in a controllable way since nearly every property of quantum gases can be perfectly controlled in the laboratory \cite{Sp}. For example, the spin-exchange interaction can be tuned by optical  \cite{Fedichev1996, Theis2004} or microwave \cite{Papoular2010} Feshbach resonance techniques. This is important since the sign of the interaction determines the ground state of spin-1 bosons, it is ferromagnetic (anti-ferromagnetic) if the interaction is negative (or positive) \cite{Ho1998, Ohmi1998, Zhang2003}.

A more powerful way to control the spinor gases is to use an external magnetic field $B$, which leads to the Zeeman effects. While the linear Zeeman effect (LZE) is relatively trivial due to the conserved total magnetic quantum number,  the quadratic Zeeman effect (QZE) resulted from the Zeeman energy difference in a spin-flip collision between different hyperfine
levels\cite{Stenger1998}, which is proportional to $B^2$,  becomes a key factor in determining the properties of the spinor BEC due to its competition with the spin-exchange interaction.  The QZE not only leads to a number of novel ground states and spin structures \cite{Stenger1998,Saito2005,Sadler2006,Murata2007,SFSP2007,KB2010,Ueda2010,UKU2010,Jacob2012}, but also
affects significantly on spin dynamics \cite{RP2004,KBB2005,Damski2007,Leslie2009,TJG2009,Mat2010,Sau2010,GJW2011,PKU2011,Ueda2011}, phase transitions \cite{Damski2007,Bookjans2011,RAK2011}, symmetry \cite{Sch2010,Ueda2010,Ueda2011,Ueda2012} and the vortex states \cite{IY2006}, and so on.

In spinor BEC,  the QZE can be induced experimentally by laser or microwave dressing field \cite{Ger2006,SFSP2007,Leslie2009,Bookjans2011}, and its sign can be varied by using either positive or negative detuning. By use of the QZE, the single-particle energies can be varied to control the dynamical
instabilities induced by the the spin mixing collisions, these instabilities provide access to a rich variety of physical
phenomena. Such controllability shows a great chance to manipulate spinor gas via QZE. Although many exciting features in spinor BEC with QZE have been investigated by both theory and experiment, a natural question is: in what extent can the QZE serve?

In this Letter, we show that the QZE can arouse inhomogeneous population exchange between the sublevels of spin-1 BEC, which leads to the pointwise different axisymmetry-breaking, and brings out the inhomogeneity of magnetization. We find two novel kinds of topologically different inhomogeneous spin domain which possess N\'{e}el-like spin domain walls determined by unmagnetized areas,  due to positive and negative QZE respectively. The spin domain can be controlled by the QZE. Furthermore, the QZE influences deeply the spin pattern and its topological structure. For the positive QZE, the surface of space-time evolution of spin density vectors is homeomorphic to the quotient space obtained from two disks by identifying two centers. This case can in some extent explain the experimental observation in \cite{Sadler2006} for a spin-1 $^{87}$Rb Bose gas. For the negative QZE, it is homeomorphic to a sphere, this case is expected to be observed in experiment. Our results provide a profound understanding for the controllability of spinor BEC via the QZE.

\textit{Spin-1 BEC with quadratic Zeeman effect--.} We start from the 3-component Gross-Pitaevskii equation in a dimensionless form
\begin{eqnarray}
&&i\partial _{t}\Phi _{\pm 1}=-\frac{1}{2}\partial _{x}^{2}\Phi _{\pm
1}+(c_{0}+c_{1})\left(|\Phi _{\pm 1}|^{2}+|\Phi _{0}|^{2}\right)\Phi _{\pm 1}\nonumber\\
&&\hspace{1cm}+(c_{0}-c_{1})|\Phi _{\mp 1}|^{2}\Phi _{\pm 1}+c_{1}\Phi
_{\mp 1}^{\ast }\Phi _{0}^{2}+(q\mp p)\Phi _{\pm 1},  \nonumber \\
&&i\partial_{t}\Phi _{0}=-\frac{1}{2}\partial _{x}^{2}\Phi
_{0}+(c_{0}+c_{1})\left(|\Phi _{1}|^{2}+|\Phi _{-1}|^{2}\right)\Phi _{0}\nonumber\\
&&\hspace{1cm}+2c_{1}\Phi _{0}^{\ast }\Phi _{1}\Phi _{-1}+c_{0}|\Phi
_{0}|^{2}\Phi _{0},  \label{eq1}
\end{eqnarray}
where $c_0$ and $c_1$ are the mean-field and the spin-exchange interactions, $p$ and $q$ correspond to the LZE and QZE respectively.
The energy is in units of $|\overline{c}_1|n$ with $n$ being the average particle density, and $\overline{c}_{1}=4\pi \hbar ^{2}(a_{2}-a_{0})/(3M)$ is the spin-exchange interaction, which is ferromagnetic if $\overline{c}_{1}<0$ (as $^{87}$Rb ) and antiferromagnetic if $\overline{c}_{1}>0$ (as $^{23}$Na). Here $a_{0}$ and $a_{2}$ denote the s-wave scattering lengths of the total spin-0 and spin-2 channels respectively \cite{Ho1998, Ueda2010}. In addition, time $t$, position $x$ and $\Phi _{\pm 1,0} $ are in units of $\hbar/(|\overline{c}_1|n)$, $\sqrt{\hbar^2/(2M|\overline{c}_1|n)}$ and $\sqrt{n}$, respectively.  For $^{87}$Rb and $^{23}$Na, the typical particle density is $n \sim 10^{14}cm^{-3}$ \cite{Stenger1998,Chang2004,Sadler2006}, thus $|\overline{c}_1|n$ is about $h \times 3.6$ Hz and $h \times 24.1$ Hz, respectively. The time and length are measured in units of $10^{-2}s $ and $10^{-6}m$, respectively.

Experimentally, the LZE can be changed independently along the condensate length by applying a field gradient $B^{\prime}$ along the axis of the trapped condensate \cite{Stenger1998}, taking as $B'z_b$ where $z_b$ is the gradient field range. The QZE can be changed independently by applying a weak external bias field $B_{0}$ \cite{Stenger1998}, which is given by $\bar{q}=\hat{q} B_{0}^2$, where $\hat{q}=\mu^2/\delta\nu_{hfs}$ ($\delta \nu_{hfs}$ is the hyperfine splitting frequency), the constant $\mu$ is defined by $\mu=g_s\mu_B/4h=700$kHzG$^{-2}$, where $g_s$ denotes the electron $g$-factor and $\mu _{B}$ is the Bohr magneton. It is found that $\hat{q}=$71.65HzG$^{-2}$ for $^{87}$Rb and 278HzG$^{-2}$ for $^{23}$Na \cite{Stenger1998}. As mentioned above, the sign and its magnitude of the QZE can also be manipulated independently by the microwave dressing field \cite{Ger2006,Leslie2009,Bookjans2011}, which provides many ways to study and/or control the spin dynamics described by the spin density vector $F=\{F_x,F_y,F_z\}$ \cite{Ueda2010}. Here $F_z$ is the longitudinal magnetization parallel to the external magnetic field, and one can define a transverse magnetization
by $|F_{\perp}|=(F_x^2+F_y^2)^{1/2}$. A state with $|F_{\perp}|\neq 0$ is called a broken-axisymmetry
phase \cite{Murata2007}. The polar angle determined by the
interaction and the magnetic field is defined by  $\theta=\arctan(|F_{%
\perp}|/F_z)$.

In the presence of the QZE, although some simplification \cite{Damski2007} and/or approximation \cite{Lamacraft2007, Black2007,SKU2007, Sau2009, Barnett2011} have been made, it seems that no analytical result has been reported. Using an ansatz with Jacobian elliptic functions, we obtain two kinds of analytical solutions corresponding to positive and negative QZE respectively.

\textit{Inhomogeneous spin domain induced by positive quadratic Zeeman effect--.} For positive QZE, we get the analytical solution of Eq. (\ref{eq1}) only under the ferromagnetic interaction ($c_1 < 0$) with the conditions $|c_1| < c_0$, which reads $\Phi _{\pm 1}=[\sqrt{-c_1/(c_0-c_1)}sn(\xi ,\lambda)\mp \sqrt{-q/c_1}dn(\xi ,\lambda )] e^{i\chi_{\pm 1}}$, $\Phi _{0}=sn(\xi,\lambda )e^{i\chi_0}$,
where $\xi =\sqrt{2q}(x-k_{1}t)$, $\lambda=\sqrt{-2c_1/(q \delta)}$ and $\chi _{\pm 1}=k_{1}x-\mu _{\pm 1}t$ with $\mu _{\pm 1}=k_{1}^{2}/2-q (c_0-2c_1)/(2 c_1)-2 c_1/\delta\mp p$,  $2\chi_0=\chi_{1}+\chi_{-1}$, $\delta=(c_0-c_1)/(c_0+c_1)$.  The requirement $|\lambda|\leq 1$ gives a restriction $q\geq -2c_1/\delta$. For $^{87}$Rb, it becomes approximately $q\geq 1.982$.

In general, the above order parameter denotes a broken-axisymmetry
phase with a periodic density distribution, the period of the wave function can be tuned via $q$. Its spin density vector reads
\begin{eqnarray}
&&F_{x}=|F_{\perp}|\cos pt, \,\, F_{y}=-|F_{\perp}|\sin pt,  \notag \\
&&F_{z}=2\sqrt{q/(c_{0}-c_{1})}sn(\xi,\lambda)dn(\xi,\lambda ),
\label{spin2}
\end{eqnarray}%
where $|F_{\perp}|=\sqrt{-8 c_1/(c_0-c_1)} sn^{2}(\xi,\lambda )$. The polar angle is  $\theta=\arctan\left(\sqrt{-\frac{2c_1}{q}}\frac{sn(\xi,\lambda)}{dn(\xi,\lambda )}\right)$.
In this case the Larmor frequency of the  magnetization is completely determined by the LZE, and the polar angle is spatial and time-dependent, which implies the inhomogeneity of the pseudospin due to the QZE.
\begin{figure}[tbp]
\setlength{\abovecaptionskip}{0pt}
\setlength{\belowcaptionskip}{-15pt}
\includegraphics[width=\columnwidth]{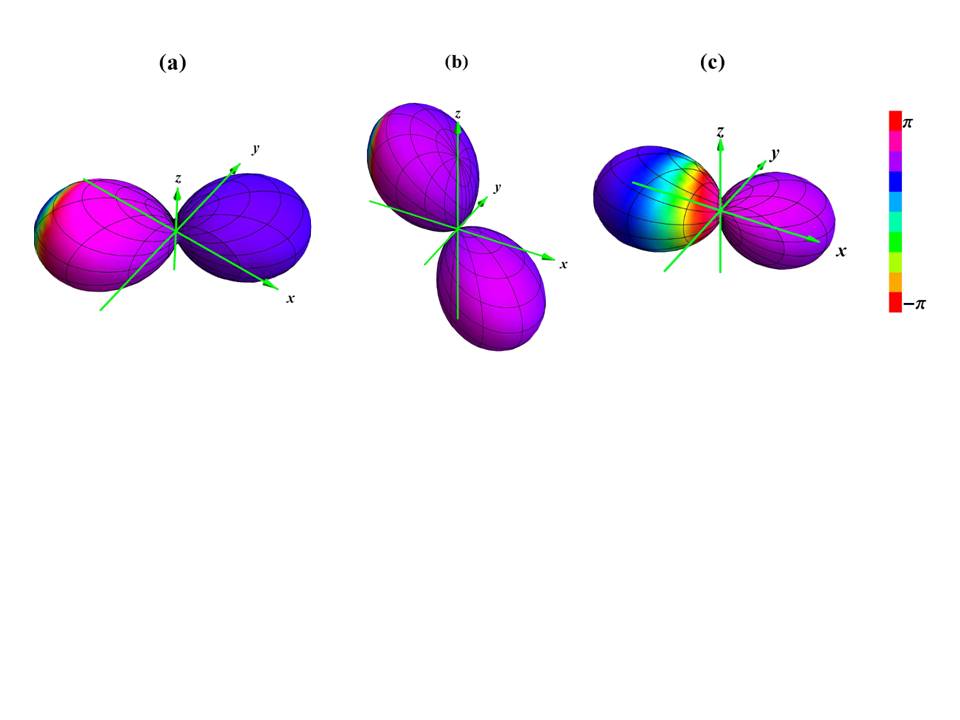}
\vspace{-3.5cm}
\caption{(Color online) Inhomogeneity of the order parameter in spin space for the solution with $q>0$. (a) and (b) show the cases for $p = 1, q = 3,x=1.7$ with different time: $t = 5.5$ for (a) and $6.5$ for (b); (c) for different magnetic field ($q = 9$) in comparison to (a).  The other parameters are taken from $^{87}$Rb atom \cite{Sadler2006} and $k_1=1$.}
\label{figm}
\end{figure}

Because an integer-spin state can be described in terms of the spherical harmonics $Y_f^m(\hat{s})$, the above order parameter can be visualized by drawing the wave functions in spin space by the surface of $|\Psi(\hat{s})|^2$, as shown in Fig. \ref{figm}, where $\Psi(\hat{s})=\sum_m \Phi_m Y_1^m(\hat{s})$, and the color represents $\arg\Psi(\hat{s})$ \cite{Ueda2010}, which is defined on $[-\pi,\pi]$. Comparing with the ground states in a uniform system \cite{Ueda2010}, Fig. \ref{figm} shows the inhomogeneity of the order parameter in spin space.  Figs. \ref{figm}(a) and (b) are different near the centers, they can not be identified via a rotation.  This implies that the QZE can cause pointwise different axisymmetry-breaking in the spin evolution,  which leads to the inhomogeneous magnetization, as described below. Fig. \ref{figm}(c) and (a) shows that the QZE can change the symmetry of the  order parameter. Furthermore,  the QZE can also change the shape near the center.

\begin{figure}[tbp]
\setlength{\abovecaptionskip}{0pt}
\setlength{\belowcaptionskip}{-15pt}
\includegraphics[width=\columnwidth]{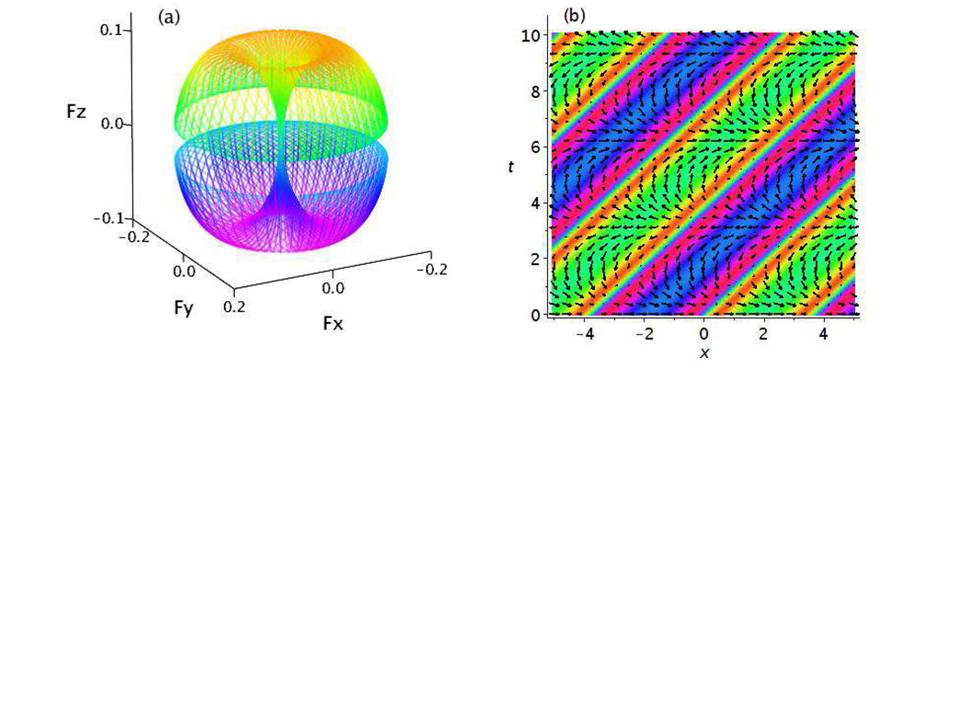}
\vspace{-3.5cm}
\caption{(Color online) Inhomogeneous magnetization described by (\ref{spin2}). (a) The spin density surface; (b) The spatial time-evolution of   $\{F_x,F_y\}$ with the density plot for $F_z$  by color as in (a). The parameters used are $p=1$, $q=2$ (for $^{87}$Rb atom \cite{Sadler2006}, $\bar{p}=10Hz$, $B^{\prime}z_b\approx 0.014mG$, $\bar{q}=20Hz$, $B_0\approx530mG$), $k_1=1$, $x\in[-5,5], t\in [0,10]$.}
\label{fig1}
\end{figure}

Eq. (\ref{spin2}) shows that along the propagation direction $\xi=const.$ such that $sn(\xi,\lambda )=0$, $F$ becomes zero periodically, which gives the unmagnetized area. When $q \rightarrow -2c_1/\delta$, i.e., $\lambda\rightarrow 1 $ (For $^{87}$Rb, $q\rightarrow 1.982$, correspondingly $\bar{q}\rightarrow 20Hz$, $B_0 \rightarrow 526$mG), the spatial and time evolution of $F$ forms a surface that looks like a sphere,  whose south- and north-poles merge into a singular point, corresponding to the zero point of magnetization. It is homeomorphic to the quotient space obtained from two disks by identifying two centers, as shown in Fig. \ref{fig1}(a). With increasing $q$ beyond $-2c_1/\delta$, two pieces of the spin surface separate gradually.  Fig. \ref{fig1}(b) shows the time evolution of the transverse magnetization vector $F_\perp=\{F_x, F_y\}$, where the lengths of the arrows are $|F_\perp|$,  the spin $z$-component density corresponds to the stripe in the same figure. Obviously, the magnetization is inhomogeneous. The heavy yellow and red regions correspond to the neighbor of north pole and south pole respectively, in such regions, $F_z$ reaches its maximum value. The green and blue lines between the yellow and red regions correspond to the zeros of $F$, an unmagnetized area. The transverse ferromagnetic domain are divided by the unmagnetized area. While the stripes form the spin domains, the unmagnetized area forms the domain wall. A further analysis together with the projection vectors $\{F_x, F_z\}$ and $\{F_y, F_z\}$ confirms that the spin domain wall has the main features of N\'{e}el wall.

Recall the experiment  in \cite{Sadler2006}, in almost quasi 1-D case, nearly pure spinor BECs  were prepared in the unmagnetized $|m_F=0\rangle$ phase with a large QZE. After a rapid quench across the quantum phase transition (of ground state), the transverse ferromagnetic domains form, which are divided by narrow unmagnetized domain walls, and the topological defects with unmagnetized filled cores have been observed. In this respect, it is constructive to compare the experiment with the analytical solution,  it is easy to conclude that near $q = -2c_1/\delta$, the spin density distribution looks like dark soliton, which indicates exactly the topological defects with unmagnetized filled cores, the experimental observation is in agreement with the features of the solution.

\begin{figure}[tbp]
\setlength{\abovecaptionskip}{0pt}
\setlength{\belowcaptionskip}{-15pt}
\includegraphics[width=\columnwidth]{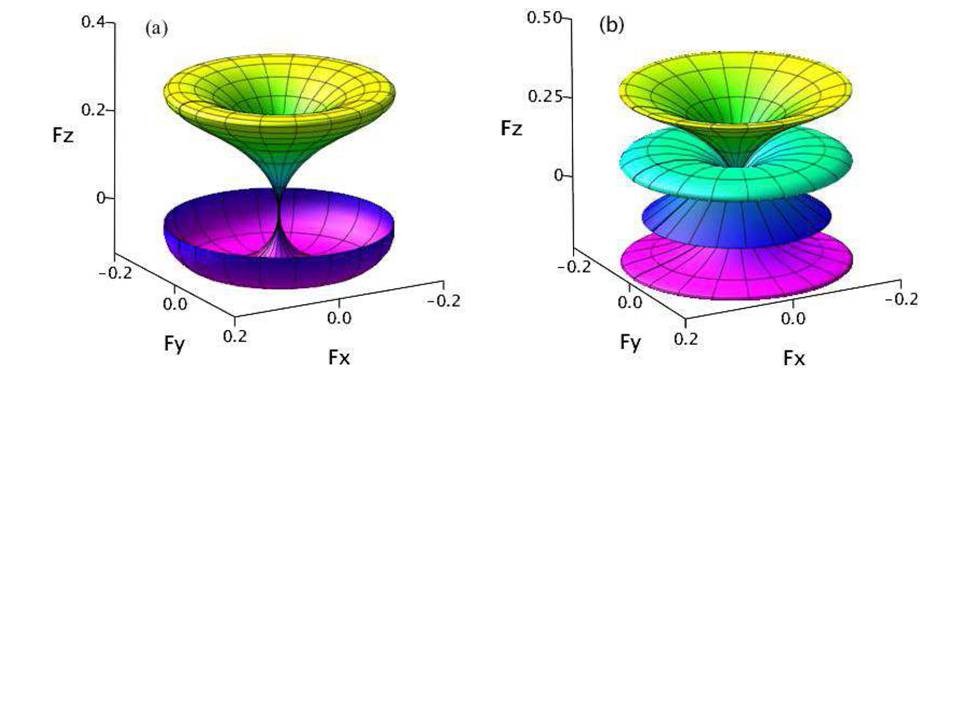}
\vspace{-3.5cm}
\caption{(Color online) Magnetization evolution in $p, q$ of (\ref{spin2}) at $t=1$ for (a) $x=2$; (b) $x=3.5$. The parameters are the same as in Fig. \ref{fig1} and $p\in [0,2\pi]$, $q\in [1.98,10]$. For $^{87}$Rb, they correspond to $\bar{p}\in [0,63Hz]$ and $\bar{q} \in [20Hz,100Hz]$ or $B^{\prime}z_b \in [0, 0.1mG]$ and $B_0\in[0.526G, 1.2G]$. }
\label{fig2}
\end{figure}

Equation (\ref{spin2}) also shows that the population exchange between the sublevels of $m_F=\pm 1,\,0$, which arises only
from the QZE, is inhomogeneous. However, it is found that in the propagation direction $\xi=0$ the system has no population exchange. This unique property corresponds to the singular point ($F=0$) of the spin density surface. In Fig. \ref{fig2} we present pointwise magnetization as a function of the LZE and QZE, which shows complicated and interesting behaviors.

When $q=-2c_1/\delta$, this solution gives a novel superposition soliton-like form for $\Phi_{\pm1}$ and a kink soliton for $\Phi_0$,
$\Phi _{\pm 1}=[\sqrt{-c_1/(c_0-c_1)}\tanh(\xi)\mp \sqrt{2/\delta}\sech%
(\xi)] e^{i\chi_{\pm 1}}$,
$\Phi _{0}=\tanh(\xi)e^{i\chi_0}$.
The shape of such solutions are determined by the ratio $\gamma=|c_0/c_1|$. For $^{87}$Rb, $\gamma
\gg 1$, each of the $\Phi_{\pm1}$ components behaves like a bright soliton. In this limit, the atom exchange between the
sublevels is strong and thus forms two domains near the soliton
propagation direction. On the contrary, far away from the direction, the atom exchange of the component $m=\pm 1$ is almost uniform,
thus no domain can further form. As such, only one spin domain
wall arises in this case.

\textit{Inhomogeneous spin domain induced by negative quadratic Zeeman effect--.} In this case, the analytical solution can be obtained for both ferromagnetic and antiferromagnetic interactions under the conditions $|c_1| < c_0$. It is known that $q>0$ for spin-1 $^{23}$Na and $^{87}$Rb BEC under a bias field.  However, as mentioned previously, the sign and magnitude of $q$ can be manipulated by using the AC Stark shift induced by using linearly polarized microwaves, therefore $q < 0$ is still accessible in experiments. Set $\nu =-2q/( c_{0}-c_{1})$, Eqs. (\ref{eq1}) has the solution:
$\Phi_{\pm 1}=A_{\pm 1}sn(\xi ,\lambda )e^{i\chi _{\pm 1}},\,\Phi_{0}=A_0 cn(\xi ,\lambda )e^{i\chi_0}$
where $A_0=\sqrt{-2A_{1}A_{-1}},\,\lambda = \sqrt{(c_{0}+c_{1})\nu}/k $, and $\xi =k( x-k_{1}t)$, $\chi
_{\pm 1}=k_{1}x-\mu _{\pm 1}t$,  $2\chi_0=\chi_{1}+\chi_{-1}$,  $\mu _{\pm 1} =
(k_{1}^{2}+k^2)/2+2c_{0}A_{-1}^{2}-2\sqrt{\nu}(c_{0}\pm c_{1})A_{-1}\pm
c_1\nu \mp p$ for given $k,\,k_1$, and the coefficients satisfy $A_1 + A_{-1} = \sqrt{\nu}$ and $%
A_1 A_{-1} \leq 0$.   We have the restriction $0\geq q \geq-A_{-1}^2(c_0-c_1)/2$. In this solution, while $q=0$ gives a periodic polar state,  $q=- A_{-1}^2(c_0-c_1)/2$ (if $A_1=0$) presents a ferromagnetic phase. For a median $q$, this solution indicates a broken-axisymmetry phase.
This suggests that by tuning $q$, it is possible to get polar phase or ferromagnetic phase
from the broken-axisymmetry phase. In this case,  the population transfer between different components is determined by the QZE, tuning $q$ will lead to the population transfer. In particular, for $k=\sqrt{(c_{0}+c_{1})\nu}$ ($\lambda = 1$), the populations of all three components can be obtained analytically\cite{note1}, from which two kinds of population transfer are identified: one is the population exchange between sublevel $m=\pm 1$ and their backgrounds, the other is the exchange between sublevels $m=0$ and $m=\pm 1$.

When both $A_{\pm1}\neq 0$, the solution represents
a  broken-axisymmetry phase with spin density vector given by
\begin{eqnarray}
&&F_{x}=-2\sqrt{-A_1A_{-1}\nu}sn(\xi, \lambda )cn(\xi, \lambda)\cos\Omega t, \nonumber \\
&&F_{y}=2\sqrt{-A_1A_{-1}\nu}sn(\xi, \lambda )cn(\xi, \lambda)\sin\Omega t,\nonumber \\
&&F_{z}=(\nu-2\sqrt{\nu}A_{-1})sn^{2}(\xi, \lambda ),\label{spinv}
\end{eqnarray}
where $\Omega=(p-c_{1}\nu +2c_{1}A_{-1}\sqrt{\nu})$
gives the Larmor frequency. The polar angle
 $\theta=\arctan(\frac{2\sqrt{-A_{1}A_{-1}}|cn(\xi,\lambda)|}
{\sqrt{\nu}-2A_{-1}|sn(\xi,\lambda)|})$. Same as the solution for $q>0$, along the propagation
direction $\xi=0$, the spin density vector becomes zero periodically, which gives unmagnetized regime, and the
magnetization is inhomogeneous.

\begin{figure}[tbp]
\setlength{\abovecaptionskip}{0pt}
\setlength{\belowcaptionskip}{-10pt}
\includegraphics[width=\columnwidth]{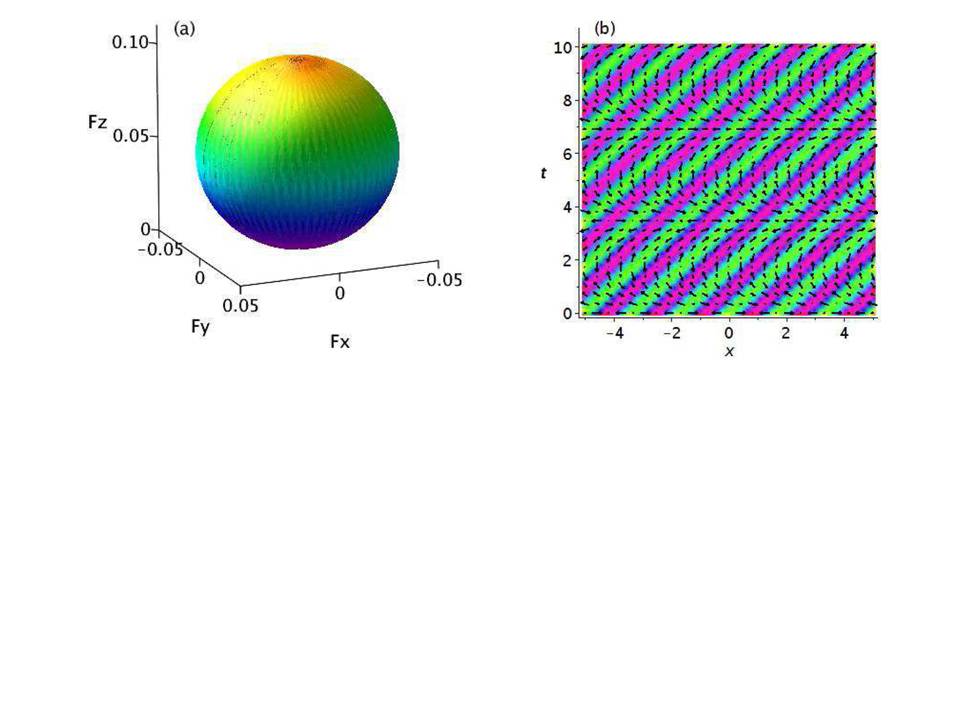}
\vspace{-3.5cm}
\caption{(Color online) Inhomogeneous magnetization described by Eq.(\ref{spinv}). (a) The spin density surface; (b) The time-evolution of the transverse magnetization vector $\{F_x,F_y\}$ with the density plot for $F_z$  by color as in (a). The parameter used are $p=1$, $q=-0.47$ (for $^{23}$Na atom \cite{Bookjans2011}, $\bar{p}=25Hz$, $B^{\prime}z_b\approx 0.035mG$, $\bar{q}=-11.6Hz$), $k_1 = 1$, $A_{-1}=-0.2$, $k=2$.}
\label{fig3}
\end{figure}

Taking $^{23}$Na as an example, Fig. \ref{fig3}(a) gives the spin density surface for $x\in [-5,5]$, $t\in [0,10]$ in which the south pole of the ellipsoid is the zero of $F$ at which the magnetization vanishes. In Fig. \ref{fig3}(b), the stripes form the spin domains and the transverse ferromagnetic domains.  A similar analysis as for $q>0$, the formed spin domain wall is also a N\'{e}el-like wall.

\begin{figure}[tbp]
\setlength{\abovecaptionskip}{0pt}
\setlength{\belowcaptionskip}{-10pt}
\includegraphics[width=\columnwidth]{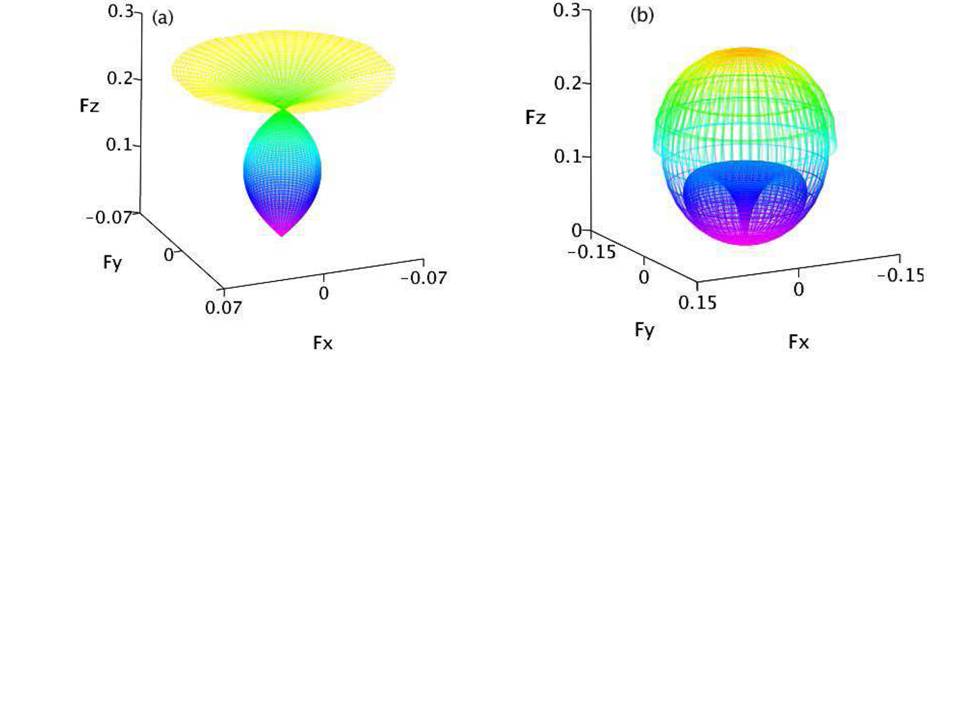}
\vspace{-3.5cm}
\caption{(Color online) Magnetization evolution in $p, q$ of Eq.(\ref{spinv}) for different positions at $t=1$.  (a) $x=2$; (b) $x=3.3$. Here $p\in[0,10], q\in[-1.88,0]$, which correspond to $\bar{p}\in[0, 247Hz]$ ($B^{\prime}z_b\approx0.35mG$ ) and $\bar{q}\in[-46.5Hz, 0]$. The other parameters are same as in Fig. \ref{fig3}.}
\label{fig4}
\end{figure}

In Fig. \ref{fig3}(b), while the red regions correspond to points near the north pole, the yellow lines correspond to the south pole where spin densities of all three components trend to zero. Other than the case of $q>0$, a spin domain is divided into two transverse ferromagnetic domains, with the domain wall being the center line of the red region, where $|F_{\perp}|=0$ but $|F_{z}|$ reaches its maximum value. In the two sides of the domain walls, the magnetization direction is opposite. However, when $q = -k^2\delta/2$,  the solution becomes  $\Phi_{\pm 1}=A_{\pm 1}\tanh \xi e^{i\chi_{\pm 1}}$, $\Phi _{0}=A_0\sech\xi e^{i\chi_0}$. The spin density surface still looks like an ellipsoid, but $|F_{\perp}|$ behaves like a split bright 1-soliton, there are two separated transverse ferromagnetic domains only concentrating  around two lines corresponding to the splitting peaks of the soliton. The topological structure of the spin configuration in $p,q$ given by Eq. (\ref{spinv}) is different from that of Eq. (\ref{spin2}), as shown in Fig. \ref{fig4}.

In the recent experiment \cite{Bookjans2011},  by rapidly switched the QZE from positive to negative values, the dynamics of an antiferromagnetic sodium BEC quenched across a quantum phase transition has been observed.  Here we point out that, if preparing  pure spinor BECs in the $|m_F=0\rangle$ phase at a large negative QZE, by the same technique as in \cite{Bookjans2011}, after a rapid quench but not going across $q=0$, the inhomogeneous  magnetization given by the solution could be observed by probing the Larmor precession \cite{Higbie2005}.

\textit{Conclusion--.} Based on analytical solutions, we have displayed  two kinds of inhomogeneous spin domain possessing N\'{e}el-like domain walls in spin-1 BEC  in the presence of positive and negative QZE respectively. Some novel inhomogeneous phenomena, which are due to the pointwise different axisymmetry-breaking arising uniquely from the QZE,  have been shown.  We present detailed description for the inhomogeneous spin pattern and the topological structures.  The transverse magnetization with positive QZE is found to be consistent with the experimental observation. The inhomogeneous magnetization for negative QZE  is expected to be observed in experiment using the method in \cite{Bookjans2011,Higbie2005}. These results display some novel phenomena in quantum magnetism, and provide a new prospect for manipulation of spinor BEC via the QZE. Our discussion is possible to be extended to spin-2 BEC case.

\textit{Acknowledgment--.} This work is supported by the FRFCU, the NKBRSFC
under grants Nos. 2011CB921502, 2010CB922904, and NSFC under grants Nos.
10934010, 11174115, and 10934008,the KPCME (No. 2011015) and
the HITSP of Hebei Province of China (No. CPRC014).


\begin{thebibliography}{99}
\bibitem{note1} In the case of $\lambda = 1$, the population of each component can be obtained analytically
\begin{eqnarray}
&&N_{1}=-\frac{\sqrt{2\delta}}{\sqrt{-q}} A_{-1}^2-\frac{4A_{-1}}{\sqrt{%
c_0+c_1}} -\frac{2\sqrt{\nu}}{\sqrt{c_0+c_1}},  \notag \\
&&N_{0}=\frac{2\sqrt{2\delta}}{\sqrt{-q}} A_{-1}^2+\frac{4A_{-1}}{\sqrt{%
c_0+c_1}}, \, N_{-1} =-\frac{\sqrt{2\delta}}{\sqrt{-q}} A_{-1}^2. \nonumber
\end{eqnarray}
Here we have made a convention that for any density function with finite
background, the atom number is defined as the number after subtracting the
background fluid contribution.  When $q\rightarrow-A_{-1}^2(c_0-c_1)/2$, we find $N_1\rightarrow 0$, $N_0\rightarrow 0$ and $N_{-1}\rightarrow 2A_{-1}/\sqrt{c_{0}+c_{1}}$.
\end{thebibliography}
\end{document}